\documentclass[letter,linenumber]{aa}  
\usepackage[switch]{lineno}
\usepackage{graphicx}
\usepackage{txfonts}
\usepackage[colorlinks=True,citecolor=blue,linkcolor=blue,urlcolor=blue]{hyperref}
\usepackage{multirow}
\usepackage{orcidlink}

\newcommand{\oiii}{\hbox{[O\,{\scriptsize III}]}}

\newcommand{\Hb}{\textrm{H}\ensuremath{\beta}\xspace}

\newcommand{\OIII}{[\textrm{O}~\textsc{iii}]\xspace}

\titlerunning{The empirical spectral resolution of NIRSpec PRISM}
\authorrunning{Zhou et al.}
\date{}
\title{An empirical calibration of the spectral resolution of JWST NIRSpec PRISM}

\begin{document} 

\author{
Hang Zhou\orcidlink{0009-0004-7133-9375}\inst{1}
\and
Xin Wang\orcidlink{0000-0002-9373-3865}\inst{1,2,3} \fnmsep\thanks{Corresponding author. \email{xwang@ucas.ac.cn}}
\and
Yuxuan Pang\orcidlink{0009-0005-3823-9302}\inst{1}
\and
Shengzhe Wang\orcidlink{0009-0007-6655-366X}\inst{1,2}
\and
Karl Glazebrook\orcidlink{0000-0002-3254-9044}\inst{4}
\and
Themiya Nanayakkara\orcidlink{0000-0003-2804-0648}\inst{5}
\and
Hu Zhan\orcidlink{0000-0003-1718-6481}\inst{2,6}
}

\institute{
School of Astronomy and Space Science, University of Chinese Academy of Sciences (UCAS), Beijing 100049, China
\and
National Astronomical Observatories, Chinese Academy of Sciences, Beijing 100101, China
\and
Institute for Frontiers in Astronomy and Astrophysics, Beijing Normal University, Beijing 102206, China
\and
Centre for Astrophysics and Supercomputing, Swinburne University of Technology, PO Box 218, Hawthorn, VIC 3122, Australia
\and
Kavli Institute for Astronomy and Astrophysics, Peking University, Beijing 100871, China
\and
Sydney Institute for Astronomy, School of Physics, The University of Sydney, Sydney, NSW 2006, Australia
}

\abstract{
We present an empirical calibration of the effective spectral resolution of the JWST NIRSpec PRISM disperser based on a large sample of galaxies drawn from archival observations. Our sample spans a wide redshift range, $0.5 \lesssim z \lesssim 10$, enabling a wavelength-dependent characterization of the PRISM resolving power across the full bandpass using the \OIII~$\lambda\lambda4959,5007$ doublets. To ensure robust measurements, we impose strict signal-to-noise requirements on both continuum and emission-line detections. Emission-line profiles are fitted with multi-Gaussian models, and the full width at half maximum (FWHM) is measured for each source. The derived empirical resolution curve shows a systematic offset from the nominal pre-launch estimate, lying between the tabulated instrumental resolution and the point-source shutter limit. We also discuss how source size affects the final spectral resolution, and provide an analytical formula that combines wavelength and morphology dependence in a root-sum-square framework. This work delivers a population-averaged, statistically robust calibration of the effective PRISM spectral resolution for typical galaxy observations. The resulting calibration is directly applicable to statistical analyses of galaxy spectra obtained with NIRSpec PRISM. Our empirical calibration exceeds the pre-launch JDox values by a factor of 1.03 to 1.63, depending on source morphology, and provides the first wavelength and morphology dependent calibration of the NIRSpec PRISM spectral resolution.
}

\keywords{Methods: observational --  Techniques: spectroscopic -- Methods: data analysis
}

\maketitle

\section{Introduction}

The Near Infrared Spectrograph (NIRSpec) onboard the JWST has been extensively used to obtain high quality near infrared spectra for a wide range of astrophysical targets. NIRSpec provides multiple observing modes, including an integral field unit (IFU) mode that delivers spatially resolved spectroscopy, and a multi-shutter assembly (MSA) mode that enables slit spectroscopy for large samples of sources. These observing modes can be combined with different dispersive elements, offering different spectral resolutions tailored to diverse scientific goals.

Among these configurations, the NIRSpec PRISM disperser is particularly powerful due to its broad wavelength coverage and high sensitivity. PRISM observations have enabled a wide range of studies on galaxy evolution and star formation across cosmic time, including measurements of gas-phase metallicity gradients \citep{jades_metallicity}, the identification of compact accreting systems such as little red dots \citep{lrd1} and the confirmation of polycyclic aromatic hydrocarbon (PAH) emission in distant galaxies \citep{pah1}.

Despite its broad scientific applications, an open question remains regarding the effective spectral resolution of JWST NIRSpec in orbit. Early analyses of NIRSpec data revealed that the measured spectral resolution deviates from the pre-launch estimates provided in the JWST User Documentation (JDox; \citealt{Jdox}), with empirical studies suggesting a systematic offset at the level of $\sim$20 - 30\%.  As a result, several early JWST studies adopted an empirical scaling factor of $\sim$1.2 - 1.3 to calibrate the spectral resolution when interpreting NIRSpec observations \citep{msafit}.

This discrepancy was recognized by the JWST instrument team, motivating dedicated in-orbit calibration efforts.  In particular, the calibration program CAL-1492 (PI: T. Beck) was designed to calibrate the NIRSpec spectral resolution using observations of an unresolved emission line source (planetary nebulae, PNe).  The analysis of these data by \cite{Shajib_2025} provided an important verification of the in-orbit performance of NIRSpec. However, the calibration based on PNe has a more limited applicability for the NIRSpec PRISM disperser. Their spectra typically contain only a small number of narrow emission lines within the PRISM wavelength coverage. As a result, such observations are less suitable for providing a fully wavelength-dependent calibration of the PRISM spectral resolution across the entire bandpass.

In addition to empirical studies, a theoretical framework for modeling the effective spectral resolution of JWST NIRSpec multi-object spectroscopy (MOS) mode has been developed by \cite{msafit}.  That work introduced the {\tt MSAFIT} software, which forward-models the spectral line spread function (LSF) by accounting for the illumination of the NIRSpec shutters and the intrinsic morphology of the source, providing theoretical predictions for the effective resolution under different observing conditions.

In this Letter, our calibration is based on direct measurements of the  $\OIII\lambda\lambda4959, 5007$ doublet from a large sample of JWST NIRSpec PRISM spectra. This bright doublet, detectable across a wide redshift range and free from absorption, provides a robust probe of the effective PRISM resolution across the full bandpass.

This paper is organized as follows. Section~\ref{sec:observation} describes the data sets used in this work. Section~\ref{sec::fit} presents the emission line fitting procedure and the measurement of the line FWHM. Section~\ref{sec::morophlolgy} introduces our empirical resolution calibration as a joint function of wavelength and source morphology. Section~\ref{sec:conclusion} summarizes the main results.

This paper focuses exclusively on the JWST NIRSpec PRISM disperser, while the resolution calibration for other NIRSpec modes falls outside the scope of this work.

\section{Observations}
\label{sec:observation}
\subsection{Samples Selection \&  Data Reduction}

To calibrate the spectral resolution of JWST NIRSpec PRISM, we construct a large galaxy sample. The sample is drawn from the DAWN JWST Archive (DJA; \citealt{brammer_2025_15472354}), which provides uniformly reduced spectra for a wide range of JWST NIRSpec observations.

To ensure reliable emission line measurements, we apply a set of selection criteria. First, we require a continuum signal-to-noise ratio (S/N) of $\mathrm{S/N}_{median} > 5$ to exclude spectra with poorly constrained continua that could bias the line profile fitting. Second, we restrict the redshift range to $z \in [0.5, 10]$. This redshift coverage ensures that the \OIII$\lambda\lambda4959, 5007$ emission line is observable across the full wavelength range of the PRISM disperser, enabling a wavelength-dependent calibration. Finally, we impose a stringent requirement on the \OIII$\lambda5007$ emission line, selecting only sources with an \OIII\ doublet S/N greater than 20 to ensure robust measurements of the line width.

\begin{figure*}
    \centering
    \includegraphics[width=0.95\linewidth]{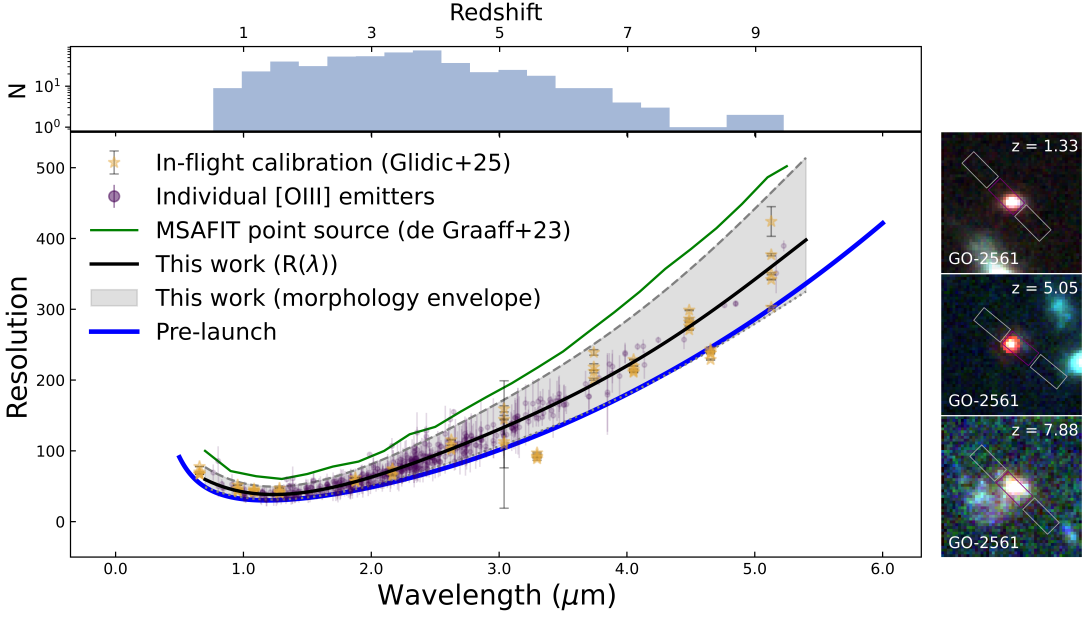}
    \vspace*{-1em}
    \caption{
    Comparison of our empirically calibrated resolution curve with other estimates for NIRSpec PRISM.
    In the lower left panel, the blue curve shows the official pre-launch resolution from JDox, measured using a uniformly illuminated slit. The gray shaded region illustrates the range of effective spectral resolutions predicted by our empirical calibration across the full range of source morphologies in our sample, from the point-source limit (upper boundary) to the largest measured source extent (lower boundary). The green curve indicates the theoretical point-source resolution predicted by {\tt MSAFIT} \citep{msafit}. The yellow points represent in-flight calibration measurements of unresolved sources from PID~6642 \citep{Glidic2025}. The black solid line denotes the $\mathcal{R}(\lambda)$ curve fitted to the full sample. The upper left panel shows the redshift distribution of our \OIII emitters analyzed here.  The right panel shows NIRCam image cutouts of the representative samples, with the position and orientation of the NIRSpec micro-shutters overlaid (rectangles). The archival JWST programs contributing to this dataset are listed in Table.~\ref{tab:jwst_surveys}.
    \label{fig:resolution}}
\end{figure*}

The final sample includes NIRSpec PRISM observations from a wide range of JWST programs, including JADES \citep{jades}, CEERS \citep{ceers}, GLASS-JWST \citep{glass-jwst}, UNCOVER \citep{UNCOVER}, WIDE MOS \citep{wide_mos} and several additional observing programs listed in Table.~\ref{tab:jwst_surveys}. The final selected sample is shown in Fig.~\ref{fig:resolution}. The sample is drawn from multiple independent JWST observing programs in the DJA compilation, indicating that our samples is not dominated by any single program or field.

For the NIRSpec PRISM disperser, the spectra are reduced using the {\tt MSAEXP} pipeline \citep{msaexp}. This pipeline performs background subtraction, wavelength calibration, flux calibration, and optimal spectral extraction, yielding one dimensional spectra for emission line fitting analysis.

\section{Emission Line Fitting}
\label{sec::fit}
We first adopt the continuum spectra and redshift estimates provided by the {\tt MSAEXP} pipeline. In {\tt MSAEXP} software, the redshift, continuum, and emission line fluxes are derived through a rapid fitting procedure based on {\tt EAZY} template libraries \citep{eazy}, combined with Chebyshev polynomials to model the continuum shape.

However, the emission line fitting implemented in {\tt MSAEXP} is primarily optimized for redshift estimation and emission line flux measurements, rather than for providing precise constraints on the velocity dispersion of individual emission lines. Therefore, in this work we refit the emission lines independently using the {\tt LMFIT} package \citep{lmfit}, while the continuum spectra are adopted from the {\tt MSAEXP}.

In the {\tt LMFIT} fitting procedure, we construct a composite emission line model consisting of three Gaussian components corresponding to \Hb, \OIII$\lambda4959$, and \OIII$\lambda5007$. The initial Gaussian widths are set according to the wavelength dependent instrumental spectral resolution of the NIRSpec PRISM dispersion solution with a single common Gaussian width adopted for all three emission lines. Initial line amplitudes are estimated from the local flux density near the expected line centers. We restrict the fitting to the \Hb, \OIII$\lambda4959$, and \OIII$\lambda5007$ transitions, as these lines form the basis of the subsequent resolution calibration.

\begin{figure}
    \centering
    \includegraphics[width=0.85\linewidth]{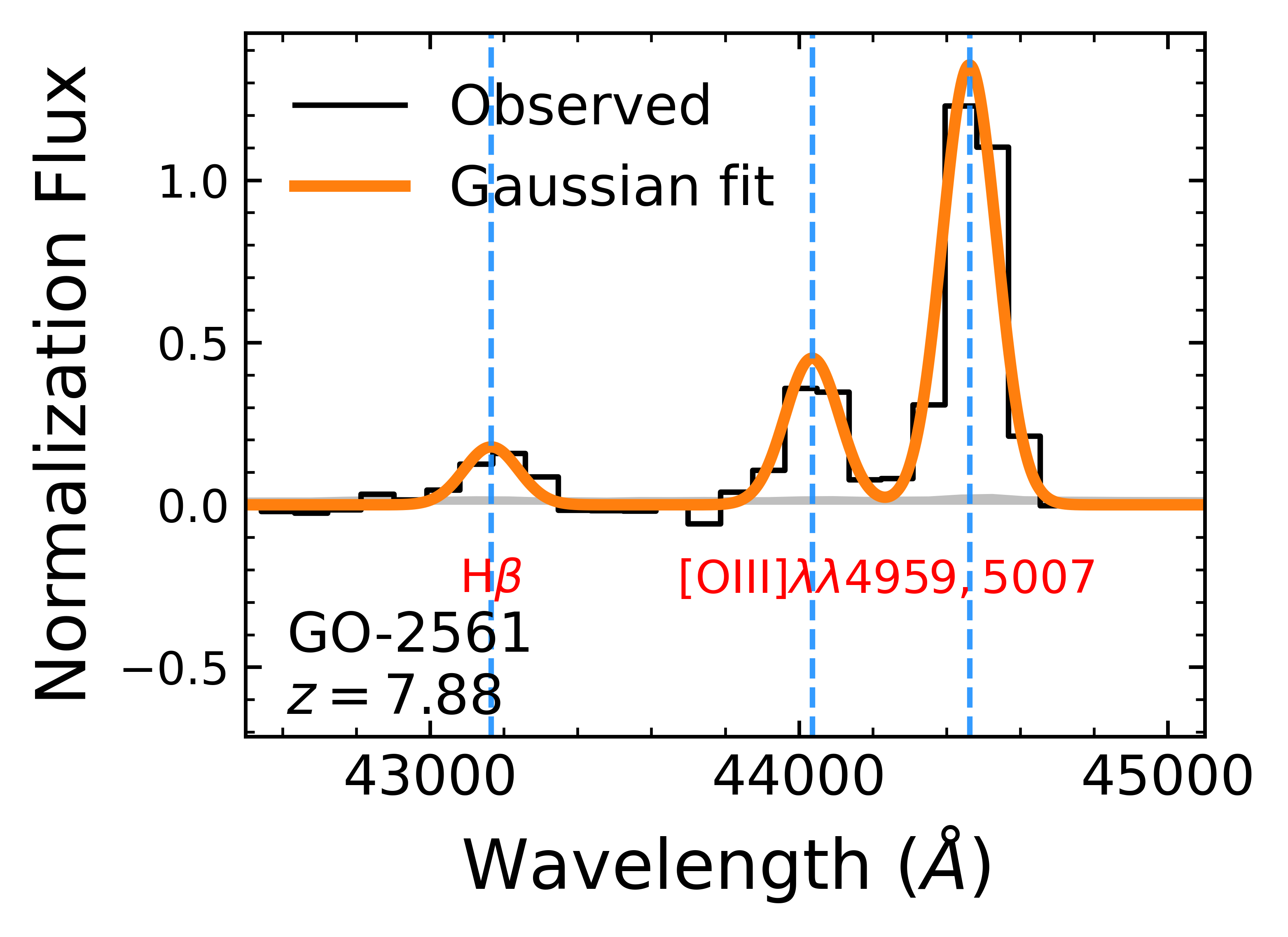}
    \vspace*{-1em}
    \caption{Example of the emission-line fitting performed in this work.
    The black step histogram shows the observed spectrum, with the gray shaded region indicating the 1-$\sigma$ uncertainty.
    The orange solid curve represents the best-fit Gaussian models, simultaneously fitting three emission lines (\Hb and \OIII~$\lambda\lambda4959,5007$) at $z = 7.88$.}
    \label{fig:fit}
\end{figure}

In the fitting procedure, we fix the intrinsic FWHM of the galaxy emission lines to 100~km~s$^{-1}$ \citep{ubler2019}. This value is adopted as a conservative upper limit on the typical intrinsic velocity dispersion of the galaxies in our sample. An example of the emission line fitting is shown in Fig.~\ref{fig:fit}.

For the \OIII doublet, we fix the flux ratio between \OIII$\lambda$4959 and \OIII$\lambda$5007 to 1:2.98. The line widths of the three emission lines are tied together so that they share the same width during the fitting.

We note that the spectra used in this work are extracted with the {\tt MSAEXP} pipeline, whose wavelength sampling differs from that of the standard JWST STScI pipeline \citep{jwst2025}. We have verified in Appendix~\ref{appendix::wave} that this difference introduces a negligible effect on the measured line widths.

\section{Spectral Resolution Calibration}
\label{sec::morophlolgy}

To characterize the empirical spectral resolution of the NIRSpec PRISM, we perform a fifth-order polynomial fit to the full sample. To further ensure robustness, we require a signal-to-noise ratio $\mathrm{S/N} > 3$ on the measured line width ($\sigma$), excluding extreme or unreliable measurements that could otherwise influence the fit.

However, a single wavelength-dependent function is insufficient to describe the full sample, because the effective spectral resolution also depends on source morphology: more extended sources illuminate a larger fraction of the shutter and broaden the LSF. To incorporate this dependence, we follow the formalism of \citet{lerner1988optics}, in which the total line broadening is described as the root-sum-square (RSS) of the individual contributions, $\mathrm{FWHM}_\mathrm{total}^{2} = \mathrm{FWHM}_\mathrm{slit}^{2} + \mathrm{FWHM}_\mathrm{resolution}^{2} + \mathrm{FWHM}_\mathrm{line}^{2}$, where $\mathrm{FWHM}_\mathrm{slit}$ arises from the convolution of the entrance slit and exit slit functions, $\mathrm{FWHM}_\mathrm{resolution}$ is the intrinsic instrumental limit set by aberrations and diffraction, and $\mathrm{FWHM}_\mathrm{line}$ is the natural width of the spectral line itself. The NIRCam data used in this work are listed in Table~\ref{tab:jwst_imaging_programs} of Appendix~\ref{appendix::data}.

To apply this framework to NIRSpec MSA observations, we proceed as follows. The source profile entering the shutter plays the role of the entrance slit image, while the NIRSpec detector pixel along the dispersion direction plays the role of the exit slit. For each source we extract a one-dimensional dispersion-direction profile by projecting the JWST NIRCam mosaic image through the corresponding MSA shutter and integrating along the cross-dispersion direction (see Appendix~\ref{appendix::Morphology} for details). We then incorporate the exit-slit (pixel) contribution by convolving this extracted profile with a Gaussian kernel of $x_0 = 0.1\arcsec$, where $x_0$ is the NIRSpec pixel size along the dispersion direction. Although the pixel response is strictly a top-hat, we adopt a Gaussian approximation of equivalent FWHM for analytical convenience; we have verified that this introduces negligible deviation from the exact convolution across the PRISM bandpass. The FWHM of the resulting convolved profile is then taken as $\mathrm{FWHM}_\mathrm{arcsec}$.

In our implementation, the morphology-dependent broadening is described through the measured quantity $\mathrm{FWHM}_\mathrm{arcsec}$, while the detector-pixel scale $x_{0}$ is retained explicitly as the characteristic angular scale of the spectrograph. The empirical calibration function $c_{0}\mathcal{R}(\lambda)$ absorbs the remaining wavelength-dependent contributions, including the intrinsic instrumental resolution in the point-source limit and any residual systematics not captured by the RSS framework, which is determined from a joint fit to the full sample. Combining these terms within the RSS framework yields:

\begin{equation}
\label{eq:resolution}
\begin{gathered}
R(\lambda, \mathrm{FWHM}_\mathrm{arcsec}) \;=\; \frac{c_{0}\,\mathcal{R}(\lambda)}{\sqrt{\,1 + (\mathrm{FWHM}_\mathrm{arcsec}/x_{0})^{2}\,}}, \\
\mathcal{R}(\lambda) \;=\; 192.5 - 316.26\,\lambda + 221.79\,\lambda^{2}  - 65.21\,\lambda^{3} + 9.69\,\lambda^{4}\\ - 0.55\,\lambda^{5}
\end{gathered}
\end{equation}

where $\lambda$ is the observed wavelength in microns, $\mathcal{R}(\lambda)$ describes the wavelength dependence of the resolution, and $c_{0}$ is the only free parameter, with a best-fit value of $c_{0} = 1.826$.

As shown in Fig.~\ref{fig:resolution}, the calibrated relation approaches the point-source prediction of \citet{msafit} at small $\mathrm{FWHM}_\mathrm{arcsec}$ and approaches the JDox pre-launch curve at larger source sizes, providing a continuous description of the effective resolution as a function of source morphology. The good agreement between the model and the data demonstrates that the $\mathrm{FWHM}_\mathrm{arcsec}$ contribution is well captured by an RSS combination. Further details are provided in Appendix~\ref{appendix::Morphology}.

\begin{figure}
    \centering
    \includegraphics[width=0.85\linewidth]{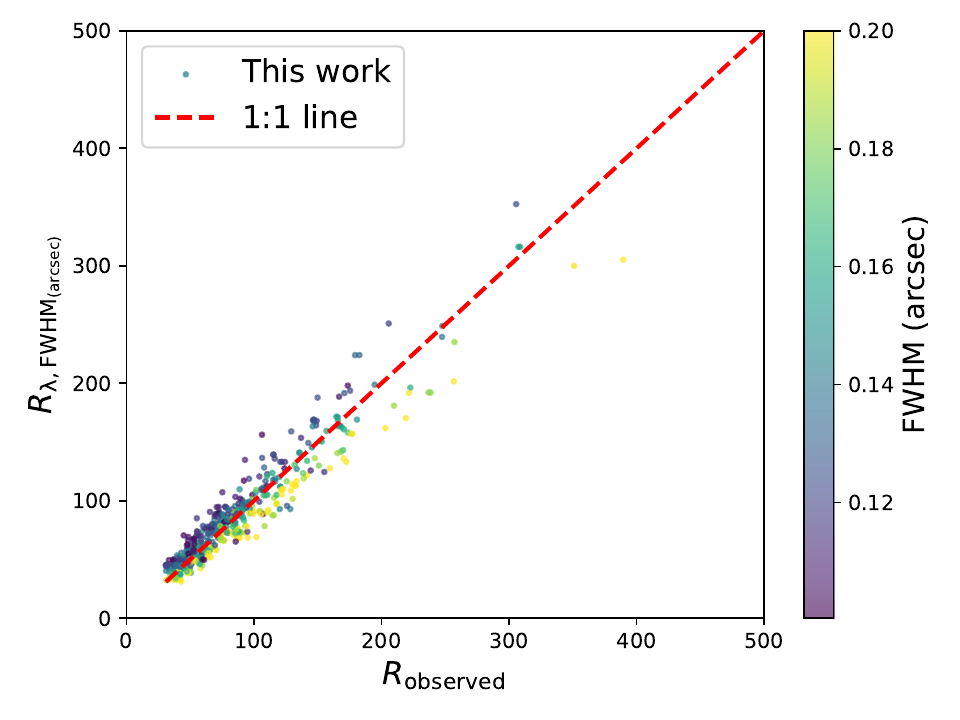}
    \vspace*{-1em}
    \caption{Comparison between the observed resolution and the model $R(\lambda, \mathrm{FWHM}_\mathrm{arcsec})$. Colored points show our fitted sample, color-coded by $\mathrm{FWHM}_\mathrm{arcsec}$, and the red dashed line indicates the 1:1 relation.}
    \label{fig:R_comparsion}
\end{figure}

Our calibration exceeds the pre-launch JDox curve by a wavelength-averaged factor of 1.03 to 1.63, depending on source morphology. Fig.~\ref{fig:R_comparsion} further compares our model with the observed data, demonstrating that the model accurately reproduces the measured resolution. We also compare our results with in-flight observations of unresolved sources from JWST NIRSpec Program ID 6642 \citep{Glidic2025}, plotted as yellow points in Fig.~\ref{fig:R_comparsion}. Our morphology envelope encompasses most of these observational data points, demonstrating that our derived resolution curve is in good agreement with the in-flight calibration measurements.  However, our sample contains relatively few point-like sources and is sparsely populated at the highest redshifts, so the calibration is less well constrained in these regimes. Despite this limitation, Eq.~\ref{eq:resolution} provides the community with a practical estimator of the effective NIRSpec PRISM spectral resolution.

\section{Summary}
\label{sec:conclusion}
In this study, we present a calibration that characterizes the effective spectral resolution of the JWST NIRSpec PRISM mode for typical galaxy observations. Our analysis reveals that the resolution of PRISM spectra exhibits an offset from the pre-launch JDox official curve \citep{Jdox}, indicating a discrepancy between the pre-launch measurements and the actual resolution achieved.

By utilizing data from DJA, we construct a large sample of high S/N \OIII emission line galaxies. The continuum is modeled with a Chebyshev polynomial, while the \OIII~$\lambda\lambda4959,5007$ doublets and \Hb lines are fitted with three Gaussian components, providing a large dataset for fitting the effective spectral resolution.

Relative to the pre-launch JDox predictions, our empirical calibration yields morphology-dependent correction factors ranging from 1.03 for the largest source sizes considered by the calibration to 1.63 in the point-source limit across the PRISM wavelength range. Our calibration therefore provides the first empirical framework for estimating the NIRSpec PRISM spectral resolution as a joint function of wavelength and morphology.

This calibration is intended to provide a population-averaged characterization of the NIRSpec PRISM spectral resolution. It can be readily applied to estimate the effective resolution of individual sources and to improve the interpretation of emission line measurements. Nevertheless, the calibration remains less well constrained in the point-source regime and at the highest redshifts because of the limited number of available objects in these regions of parameter space. Despite these limitations, our results provide a practical and conservative benchmark for the community, particularly for statistical analyses of large galaxy samples.

\begin{acknowledgements}
We thank the referee, Dr. Peter Zeidler, for very insightful comments that helped improve the quality of this paper. HZ thanks Dr. Chunyi Zhang for helpful discussions that helped improve the manuscript. 
This work is supported by the National Natural Science Foundation of China (grant 12373009), the China Manned Space Program with grant no. CMS-CSST-2025-A06, the National Key R\&D Program of China No.2025YFF0510603, the CAS Project for Young Scientists in Basic Research Grant No. YSBR-062, and the Fundamental Research Funds for the Central Universities. XW acknowledges the support by the Xiaomi Young Talents Program, and the work carried out, in part, at the Swinburne University of Technology, sponsored by the ACAMAR visiting fellowship. 
This work is based on observations made with the NASA/ESA/CSA JWST.
The data were obtained from the Mikulski Archive for Space Telescopes at the Space Telescope Science Institute, which is operated by the Association of Universities for Research in Astronomy, Inc., under NASA contract NAS 5-03127 for JWST.
\end{acknowledgements}

\bibliography{sample701}{}
\bibliographystyle{aasjournalv7}

\appendix
\section{Morphology effects on the spectral resolution}\label{appendix::Morphology}

In this section, we investigate how source morphology affects the effective spectral resolution. The key quantity of interest is the one-dimensional FWHM along the dispersion direction, which directly reflects the spatial distribution of light entering the NIRSpec MSA.

Following the method of \citet{Nanayakkara2024}, we project the two-dimensional NIRCam image along the cross-dispersion direction and measure the one-dimensional FWHM along the dispersion direction, as it is the spatial extent in this direction that primarily determines the effective spectral resolution.

To characterize the morphology-dependent contribution to the spectral resolution, we first estimate the dispersion-direction light distribution of each source. Given the similar point spread functions (PSF) of JWST NIRCam and NIRSpec, we adopt the reasonable approximation that NIRCam imaging provides a proxy for the light profile entering the NIRSpec shutter.

Using the spectroscopic redshift of each galaxy, we identify the NIRCam filter containing the redshifted \OIII$\lambda5007$ emission and extract a corresponding image cutout for subsequent analysis. 
% The NIRCam mosaics are drizzled onto a finer pixel scale, which reduces the sensitivity of the FWHM measurement to pixelization effects.

To determine the location and orientation of the corresponding NIRSpec MSA, we utilize the {\tt MSAEXP} software package. Specifically, we extract all relevant MSA configuration files and identify the shutter associated with each source, allowing us to map the shutter geometry onto the NIRCam image frame.

To accurately extract the one-dimensional dispersion profile, we project the image data along the cross-dispersion direction. The NIRCam mosaics are drizzled onto a finer pixel scale, reducing the sensitivity of the FWHM measurements to pixelization effects. To determine the precise dispersion direction, we employ a Principal Component Analysis (PCA) approach on the image cutouts. This method allows us to robustly identify the major and minor axes of the shutter geometry directly from the data, avoiding the potential instabilities and uncertainties associated with relying solely on image rotation angles provided in the headers.

To ensure a robust measurement of the source profile, particularly for low-redshift galaxies that may fill or extend beyond a single shutter, we extract the profile over a window of $0.6\arcsec$ along the dispersion direction, corresponding to the width of three shutters. This wider extraction window provides a more reliable characterization of both the profile wings and the local background than a single-shutter extraction. We additionally require the source centroid to lie within the central shutter, excluding configurations for which the source is significantly offset from the shutter center and therefore not well represented by our modeling assumptions. We then fit a Gaussian model to the full extracted profile to obtain its FWHM.

To account for the difference between the NIRCam and NIRSpec PSFs, we use {\tt WebbPSF} \citep{webbpsf} to generate the NIRCam PSF for each band and resample it onto the NIRSpec pixel size (0.1\arcsec). To capture the uncertainty introduced by sub-pixel positioning, we perform 1000 Monte Carlo realizations with random sub-pixel offsets and take the median FWHM. We then transform the measured FWHM to the NIRSpec frame in quadrature as
\begin{linenomath}
\begin{equation}
    \mathrm{FWHM} = \sqrt{\mathrm{FWHM}^{2} - \mathrm{FWHM}_\mathrm{NIRCam}^{2} + \mathrm{FWHM}_\mathrm{NIRSpec ,resampled}^{2}},
    \label{eq:psf_correction}
\end{equation}
\end{linenomath}
 Finally, we compute the length of this PSF-corrected FWHM interval that geometrically overlaps with the central shutter; this clipped length is adopted as $\mathrm{FWHM}_\mathrm{arcsec}$, representing the effective source size along the dispersion direction and the characteristic scale responsible for morphology-induced spectral broadening.

Specifically, our selection requires a signal-to-noise ratio of $S/N > 3$ for the \oiii\ emission-line fit, along with a minimum spatial width of $\mathrm{FWHM}_{\mathrm{arcsec}} \ge 0.1''$ to ensure robust measurements.

Our final sample consists of approximately 500 sources with available NIRCam imaging. To minimize contamination from AGN activity or outflow-driven line broadening, we have excluded three sources exhibiting clearly broadened line profiles, identified by cross-matching with available high-resolution grating observations from the DJA archive.

\section{Effect of Wavelength Resampling on the Measured Line Width}\label{appendix::wave}
In this appendix, we investigate how different wavelength resampling schemes may affect the measurement of emission line widths. In particular, we compare the wavelength sampling used in the {\tt MSAEXP} extraction with the wavelength grid produced by the standard JWST STScI pipeline.

To evaluate the potential impact of these different sampling methods, we perform a set of simulations. We first generate a synthetic spectrum consisting of a single Gaussian emission line in 10000 times. The central wavelength of the line is randomly selected within the range 0.6 $\mu m$ – 5.3 $\mu m$ in order to sample the full wavelength coverage of the NIRSpec prism. To accurately represent the intrinsic line profile, we construct the model spectrum on an oversampled wavelength grid that is ten times finer than the original wavelength array derived from the JWST dispersion solution. The Gaussian line width is determined from the instrumental resolution by converting the Eq.~\ref{eq:resolution} ($\mathcal{R}(\lambda)$) resolution into a Gaussian $\sigma$. In addition, we introduce an additional broadening of 100 ~km~s$^{-1}$ to mimic the intrinsic velocity dispersion of galaxies.

Next, we compare the wavelength arrays derived from the {\tt MSAEXP} pipeline and the official JWST STScI pipeline. This is particularly relevant since the spectra used in this work are drawn from the DJA data products, which employ {\tt MSAEXP} for one dimensional spectral extraction. Here, we investigate whether the choice of wavelength array, which is extracted from different pipeline, introduces any systematic differences in the empirically derived spectral resolution formula obtained in this work. The oversampled Gaussian model is then interpolated onto the two wavelength grids ({\tt MSAEXP} and the official pipeline). We subsequently fit these resampled spectra using a Gaussian model implemented with {\tt LMFIT} in order to evaluate how differences in wavelength sampling affect the recovered line widths.

To ensure that the fitting process does not trivially recover the input parameters, we intentionally offset the initial guesses of the line center and $\sigma$ from the true values. This approach avoids introducing prior knowledge of the true parameters into the fitting procedure and allows us to more realistically assess the influence of wavelength sampling on the measured results.

Our results demonstrate that due to the differences in wavelength arrays, slight offsets in the fitted values can arise during resampling. However, this effect is very minor, introducing less than 0.5\% offsets in the overall recovery of the line widths. However, at shorter wavelengths, the spectral resolution of the prism becomes relatively low. As a result, the \OIII$\lambda$4959, \OIII$\lambda$5007, and \Hb\ emission lines can become partially blended, which requires a three-component Gaussian model to properly describe the line profile.

To investigate this situation, we performed an additional simulation using a three Gaussian model consisting of the \OIII$\lambda$4959, \OIII$\lambda$5007, and \Hb\ emission lines. The simulation procedure follows the same approach described above. For each realization, we randomly select a central wavelength and repeat the process 10,000 times to construct the blended three line models.

The relative positions of the three Gaussian components are fixed according to the rest frame wavelengths of the emission lines. The amplitude ratio of the \OIII\ doublet is fixed at $1:2.98$ for \OIII$\lambda$4959\ and \OIII$\lambda$5007. The amplitude of the \Hb\ line is set equal to that of \OIII$\lambda$4959\ to represent a typical emission line galaxy. The wavelength range of the simulation is limited to 0.6$\mu m$ - 4.5$\mu m$, since at wavelengths longer than 4.5 $\mu m$ the Gaussian components become sufficiently separated and the blending between the lines is no longer significant. The fit results again show only a negligible difference, consistent with the findings above.

The NIRSpec detector slightly undersamples the point spread function at the blue end of the wavelength range. As a result, the intrinsic LSF can be narrower than the detector pixel scale. Depending on the exact subpixel position of the source, this may introduce small shifts in the measured line center and profile when the spectra are resampled. This could introduce uncertainties on the resolution measured.

These results indicate that the derived calibration is not sensitive to the choice of wavelength grid. This implies that the resampling and drizzling procedures used in the {\tt MSAEXP} extraction introduce only a negligible difference in the measured line widths. Therefore, while the calibration presented in this work is formally derived using the {\tt MSAEXP} DJA extractions, the overall impact of the wavelength grid differences between {\tt MSAEXP} and the standard STScI pipeline is expected to be minimal for practical applications.

While we adopt Gaussian profiles throughout this work, we note that subtle Lorentzian-like wings could broaden the inferred line widths by up to $\sim 4\%$ (JWST/NIRSpec team, priv. comm.), which is comparable to our typical measurement uncertainties and expected to have a relatively minor impact on our results.

Of course, in the current analysis, we only examined the difference arising from the wavelength sampling method between {\tt MSAEXP} and the standard STScI pipeline. However, the extraction method itself may also introduce non-negligible differences, and a detailed investigation of this effect would require a dedicated and systematic comparison of the two extraction methodologies, which is beyond the scope of this paper.

\section{Summary of Utilized JWST Datasets}\label{appendix::data}

In this work, we utilize archival JWST data comprising both NIRSpec spectroscopy and NIRCam imaging across multiple programs. Tables~\ref{tab:jwst_surveys} and \ref{tab:jwst_imaging_programs} summarize the NIRSpec spectroscopic and NIRCam imaging observations used in our analysis, respectively, along with their corresponding program IDs.

\begin{table}
\caption{JWST archival programs with NIRSpec PRISM data used in this work.}
\label{tab:jwst_surveys}
\centering
\begin{tabular}{lcc}
\hline
\small
\setlength{\tabcolsep}{1pt}
\renewcommand{\arraystretch}{0.8} 
Program & Proposal ID& Number of \OIII emitters \\
\hline
UNCOVER      & 2561 & 116 \\
JADES        & 1180, 1181, 1210, 1286 & 592 \\
WIDE MOS     & 1211 - 1215 & 368 \\
CAPERS       & 6368 & 68 \\
CEERS        & 1345 & 100 \\
Others       & -- & 731 \\
\hline
\end{tabular}
\tablefoot{
Other programs include those with proposal IDs:
1433, 1747, 2028, 2073, 2198, 2282, 2565, 2750, 2756, 2767,
3073, 3215, 4233, 4557, 6541, and 6585.
}
\end{table}

\begin{table}
\caption{JWST archival programs contributing to the imaging data used in this work.}
\label{tab:jwst_imaging_programs}
\centering
\small
\setlength{\tabcolsep}{8pt}
\renewcommand{\arraystretch}{1.0}
\begin{tabular}{lc}
\hline
Program & Proposal ID \\
\hline
UNCOVER & 2561 \\
UNCOVER--MegaScience & 4111 \\
JADES & 1180, 1181, 1210, 1286, 1287, 3215, 4540 \\
GLASS-JWST & 1324 \\
CEERS & 1345 \\
COSMOS-Web & 1727 \\
PRIMER & 1837 \\
FRESCO & 1895 \\
CAPERS & 6368 \\
PEARLS & 1176, 2738 \\
Blue Jay & 1810 \\
NGDEEP & 2079 \\
MIDIS & 1283, 6511 \\
PANORAMIC & 2514 \\
ALT & 3516 \\
CONGRESS & 3577 \\
BEACON & 3990 \\
POPPIES & 5398 \\
OASIS & 5997 \\
SAPPHIRES & 6434 \\
VENUS & 6882 \\
Others & -- \\
\hline
\end{tabular}
\tablefoot{Other programs have proposal IDs 1264, 1433, 1840, 2073, 2198, 2279, 2282, 2516, 2750, 2756, 2767, 3073, 3954, 4762, 6480, 6541, 6585, 8060, 9372.}
\end{table}

\end{document}